\documentclass[twocolumn,preprintnumbers,amsmath,amssymb]{revtex4}
\usepackage{graphicx}% Include figure files
\usepackage{dcolumn}% Align table columns on decimal point
\usepackage{bm}% bold math

\begin{document}
\title{\bf Strain engineered  graphene using a nanostructured substrate: II Pseudo-magnetic fields}
\author{M. Neek-Amal$^{1,2}$ and F. M. Peeters$^2$ }
\affiliation{$^1$Department of Physics, Shahid Rajaee Teacher Training University,
Lavizan, Tehran 16785-136, Iran.\\$^2$Departement Fysica,
Universiteit Antwerpen, Groenenborgerlaan 171, B-2020 Antwerpen,
Belgium.}
\date{\today}
\begin{abstract}
The strain induced pseudo-magnetic field in supported graphene
deposited on top of a nanostructured substrate is investigated by
using atomistic simulations. Step, elongated trench, one dimensional
barrier, spherical bubbles, Gaussian bump and Gaussian depression
are considered as support structures for graphene. From the obtained
optimum configurations  we found very strong induced pseudo-magnetic
fields which can reach up to  $\sim$ 1000\,T due to the
strain-induced deformations in the supported graphene. Different
magnetic confinements with controllable geometries are found by
tuning the pattern of the substrate. The resulting  induced magnetic
fields for graphene on top of a step, barrier and trench are
calculated. In contrast to the step and trench the middle part of
graphene on top of a barrier has zero pseudo-magnetic field. This
study provides a theoretical background for designing magnetic
structures in graphene by nanostructuring substrates. We found that
altering the radial symmetry of the deformation, changes the
six-fold symmetry of the induced pseudo-magnetic field.

\end{abstract}
\maketitle

\section{Introduction}
In most of the experiments on graphene, the 2D atomic layer is
placed on top of a substrate, which at atomic scale is not flat.
Geometrically structured substrates affects various properties of
graphene~\cite{geim,novo}, and can prevent the crumpling of graphene
which occurs for free standing graphene without
support~\cite{nelson}. Recently, the modification of the properties
of graphene on top of a substrate were investigated. It was found
that substrates can induce corrugations, modify the electric
conductance and deform graphene~\cite{bao,sio2}.

Tomori \emph{et al} used pillars made of a dielectric material
placed on top of a substrate which is then overlayed with graphene
to generate non-uniform strain on a micro-scale~\cite{tomori}.
Elastic deformations in graphene creates a pseudo-magnetic field
which acts on graphene's massless charge
carriers~\cite{naturegoonia,revmodphys,PRBmidgap}. The resulting
variation of the hopping energies can be viewed as an  induced
pseudo-magnetic field which enters in the Dirac equation.
Engineering the right topology of the induced pseudo-magnetic field
can provide magnetic confinement which confines electrons in
specific regions in space~\cite{peeters1999,peters2001}. It has been
shown theoretically that inhomogeneous magnetic fields are able to
confine massless Dirac fermions in a monolayer graphene
sheet~\cite{PRL20007}.

Here, we investigate several nano-structured substrates with
different geometrical deformations. We carried out molecular
dynamics simulations at $T$=300\,K to minimize the energy and find
the optimum profile of the deposited graphene on top of different
nanostructured substrates. Elongated trench, barrier, bubble,
Gaussian bump and Gaussian depression are considered as examples of
nano-structured substrates. The adhesion of the substrate to the
deposited graphene can induce a very strong pseudo-magnetic field
which we found depends on the imposed boundary conditions on the
graphene sheet. Strong pseudo-magnetic fields ($\sim$ 1000 T) are
found around the deformed regions in graphene. A substrate with: (i)
a step, forms two magnetic-barriers around the step with opposite
sign, (ii) a trench forms two narrow magnetic-barriers around the
trench boundaries with the same sign and one with opposite sign
within the trench, and (iii) a one dimensional barrier forms two
pairs of magnetic-barriers around the barrier's wall. The magnetic
confinement for a Gaussian depression in the substrate looses the
six-fold symmetry of the pseudo-magnetic field which is not the case
for  GE on top of a Gaussian bump.

This paper is organized as follows. In Sec. II the details of the
atomistic model are presented.  In Sec. III we present  the strain
induced gauge field model. In Sec.~IV we present  results for the
 gauge fields and the pseudo-magnetic fields,  for
various nano-structured substrates. The results are summarized in
Sec. V.

\section{Atomistic model}
In order to find the optimum configuration of graphene (GE) on top
of various nanostructured substrates we employed classical atomistic
molecular dynamics simulation (MD). The second generation of
Brenner's bond-order potential ~\cite{brenner2002} is employed for
carbon-carbon interaction and the van der Waals (vdW) interaction
between GE and different substrates is modeled by employing the
Lennard-Jones (LJ) potential, i.e.
\begin{equation}\label{Equr}
u(r)=4\epsilon[(\sigma/r)^{12}-(\sigma/r)^6],
\end{equation}
 where $r$ is the distance between the two particles,
$\epsilon$ and $\sigma$ are the `\emph{energy parameter}' and  the
`\emph{length parameter}', respectively. To model the interaction
between two different types of atoms such as the carbon atom (C) and
the substrate atom (S), we adjust the LJ parameters using the
equations $\epsilon_T\,=\, \sqrt{\epsilon_C \epsilon}$ and
$\sigma_T\,=(\sigma_{C}+\sigma)/2$. For carbon we use the parameters
 $\sigma_C=$3.369\,\AA~and $\epsilon_C=$2.63\,meV. For the
substrate atoms we set $\sigma=$3.5\,\AA~ and $\epsilon=$10.0\,meV,
 which is typical e.g. for a SiO$_2$ substrate~\cite{MD2010}. The simulation
 is done for a GE sheet with dimension $l_x=$19.17\,nm and
$l_y=$19.67\,nm at $T$=300\,K. The number of substrate atoms is
$M=$6000. In order to model the substrate, a (100) surface having a
typical lattice parameter $\ell$=3\,\AA ~is assumed. The density of
sites in the substrate is $\Sigma_S=\ell^{-2}$. The details of the
found deformations are reported in our previous
study~\cite{neekpeet}.
\begin{table*}
\caption{A list of all relevant parameters used in the paper.}
\begin{tabular}{| l | l | c | c |}
  \hline
$l_x,l_y$&The graphene length and width\\
$\epsilon,\sigma$&The energy and length parameters in the van der Waals (vdW) potential for the substrate atoms,  Eq.~(\ref{Equr})\\
$\lambda,\theta(x)$&The wave length and the step function\\
$R$&The radius of the Gaussian bump or depression\\
$h_0$&The amplitude of sinusoidal waves or height (depth) of Gaussian bump/bubble/barrier (depression or trench)\\
$h_1,d$&A shift or vertical distance between graphene and substrate and the width of the trench/barrier\\
$u_{\alpha \beta}$,\textbf{A},B&Strain tensor, strain induced gauge field and magnetic field\\
 \hline
\end{tabular}
\label{table1}
\end{table*}

\section{Strain induced pseudo-magnetic field}
Generalizing the Dirac equation, which governs the low energy
electronics of graphene, to curved surfaces is an interesting
development which may model some cosmological problems
~\cite{revmodphys,PRBmidgap}. The metric of the curved surface
enters now into the Dirac equation. The origin of the deformations
are external stresses which deform graphene so that the nearest
neighbor distances become non-equal. Notice that the external
stresses can be induced by the substrate. The latter results in
modified hopping parameters introduced in the tight-binding model
which are now a function of the atomic positions
$t(\textbf{r})$~\cite{PRBrapid}. Assuming small atomic displacements
(i.e. $\textbf{u}=\textbf{r}'_i-\textbf{r}_i<a_0$ where $a_0$ is the
carbon-carbon bond length) and rewriting the Dirac Hamiltonian in
the effective mass approximation with nonequal hopping parameters
tells us that the strain induces an effective  gauge field
\begin{equation}\label{gauge}
\textbf{A}=\frac{2\beta\hbar}{3a_0 e}(u_{xx}-u_{yy},-2u_{xy}),
\end{equation}
where $\beta$ ($\sim$2-3) is a constant  and $u_{\alpha\beta}$ is
the strain tensor including out of plane
displacements~\cite{revmodphys}. The corresponding pseudo-magnetic
field perpendicular to the $x-y$ plane is obtained as
\begin{equation}\label{Magnetic}
B=\partial_y A_x-\partial_x A_y.
\end{equation}
This is the  pseudo-magnetic field which the electron experiences in
the K valley. We will find $B$ by making the necessary
differentiations numerically for longitudinally supported boundary
conditions. Here we are mostly interested in the out-of-plane
contributions of  the pseudo-magnetic field which mainly appears
around the deformed parts of  graphene. The other in-plane terms
contribute less to the pseudo-magnetic field around the deformed
parts, particularly when the system is larger than the size of these
deformed parts and is supported from boundaries. Notice that in
order to perform the numerical differentiations (in  Eq.~(2) and
Eq.~(3)) one needs a reference graphene lattice ($\textbf{r}_i$) in
order to compare the optimized lattice ($\textbf{r}'_i$) with the
reference system. We used the optimized graphene profile at the
given  temperature over  a flat substrate as the reference system.
However, when the boundaries are free there is  considerable
difference (at the boundaries and for some particular systems)
between the optimized graphene over the deformed substrate and the
reference system. This is due to the fact
 that at finite temperature the free edges \emph{of graphene over the substrate
 can vibrate and deform (due to the substrate induced strain)}
  freely while they will not be deformed
 in the reference system.
Therefore the reference system with free boundaries for some of the
systems can be very different from the \emph{optimized graphene over
the deformed substrate at finite temperature}, hence the
differentiation is not well defined. Therefore, in this paper we
 focused will be on systems with fixed boundaries, which were studied in
our  previous paper~[15], where we have a true reference system
suitable for numerical differentiations.

\section{Results and discussion}
In this study we investigate several different geometries for the
substrate which can be realized experimentally. For all studied
cases we first  obtained the optimum configuration of GE on top of
the different nanostructured  substrates using MD simulations (those
results were presented in our previous work~\cite{neekpeet}). Then,
for the supported boundary condition, we calculate the corresponding
gauge field from which we obtain the pseudo-magnetic field.

\subsection{Step}
\begin{figure}
\begin{center}
\includegraphics[width=0.98\linewidth]{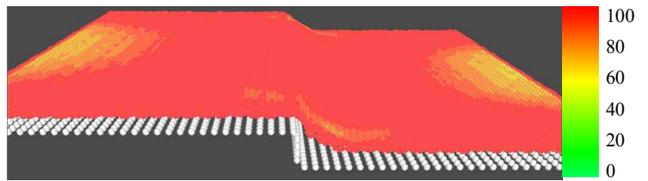}
\caption{(Color online)  The optimum configuration of arm-chair
graphene over a step located at $x=0$ with supported longitudinal
ends. The colors indicate the size of strain. \label{figstepsup} }
\end{center}
\end{figure}

An interesting substrate configuration is a step which was recently
studied in an experiment to measure the electronic and morphology of
 deposited graphene~\cite{stepPRB}
\begin{equation}\label{Eqstep}
h_S(x,y)=h_0\theta(x),
\end{equation}
where $\theta(x)$ is the Heaviside step function and $h_0=$1\,nm~is
the height of the step.  GE with arm-chair direction is put on top
of the step. In Fig.~\ref{figstepsup}  the optimum configuration of
GE along the arm-chair direction with longitudinally supported
boundary condition is shown when placed over a sharp step defined
by~Eq.~(\ref{Eqstep}).

\begin{figure}
\begin{center}
\includegraphics[width=0.8\linewidth]{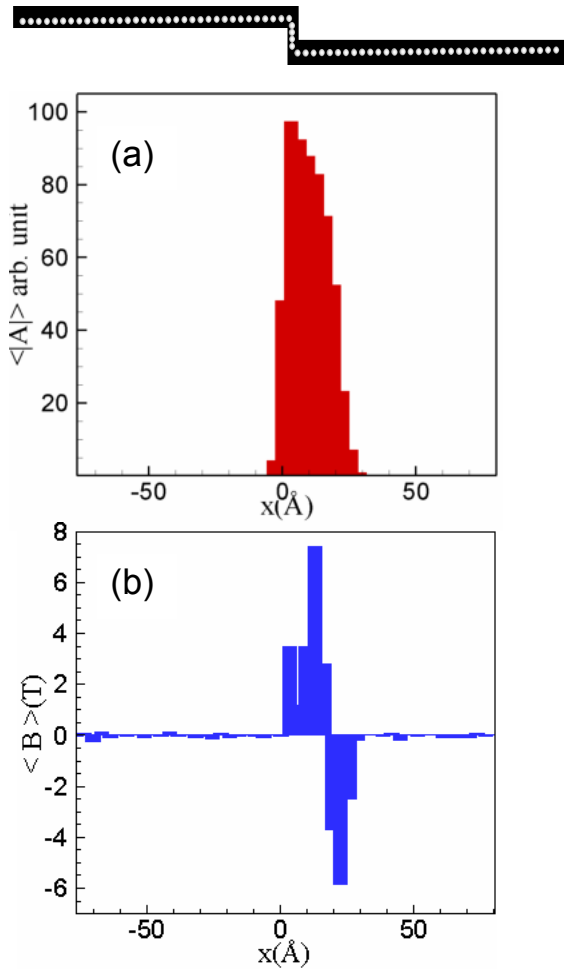}
\caption{(Color online)  The averaged   gauge fields (a)
 and  the induced pseudo-magnetic fields (b) averaged over the $y$-direction for
graphene on top of a step as shown in Fig.~\ref{figstepsup} which
has been supported from the longitudinal ends while it can freely
moves along the $z$-direction.  \label{ABstep} }
\end{center}
\end{figure}

%\begin{figure}
%\begin{center}
%%\includegraphics[width=0.3\linewidth]{ustep.eps}
%%\includegraphics[width=0.32\linewidth]{astep.eps}
%\includegraphics[width=1\linewidth]{bstep.eps}
%\caption{(Color online)  The induced pseudo-magnetic fields for
%graphene shown in Fig.~\ref{figstepsup} which is supported at the
%longitudinal ends while it can freely moves along the $z$-direction.
%\label{figstepB} }
%\end{center}
%\end{figure}

The induced gauge field  as obtained from Eq.~(\ref{gauge}) is
averaged over the $y$-direction and is shown in Fig.
\ref{ABstep}(a). All atoms at the step region  are stretched which
results in considerable gauge fields around $x\approx0$.
Fig.~\ref{ABstep}(b) shows the averaged pseudo-magnetic field over
the $y$-direction, $\langle B \rangle$, versus $x$. In order to
calculate averages we made a histogram where  $l_x$ is divided into
60 equal parts.
%figure $\langle B \rangle$s are mostly less than 10\,Tesla.
 Notice that the induced pseudo-magnetic field is mostly
concentrated beyond $x$=0 and consists of a positive and an adjacent
negative barrier with total average zero. Because of thermal
fluctuations (i.e. $T$=300\,K) the positive and negative barrier are
only approximately identical. The larger the curvature the larger
the magnetic field. The large pseudo-magnetic field around the step
separates the GE sheet into a left and a right hand side, where `B'
is small. Electrons will be trapped in this region into snake orbits
and electrons passing perpendicular to this rectangular part will
experience large pseudo-magnetic fields. Notice that by changing the
height of the step ($h_0$), we are able to control the size of the
magnetic barrier and consequently the magnetic confinement.

\subsection{ Trench}

\begin{figure}
\begin{center}
\includegraphics[width=0.98\linewidth]{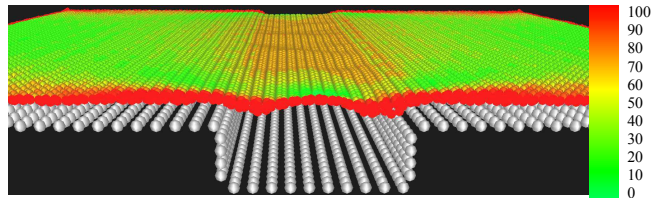}
\caption{(Color online)  The optimum configuration of arm-chair
graphene over a trench located at $|x|<1.5$\,nm where both
longitudinal ends were supported in the $x-y$ plane. The colors
indicate the size of the strain. \label{figwellsup} }
\end{center}
\end{figure}

The other important  substrate that we study here is an elongated
trench
\begin{equation}\label{Eq.trench}
 h_S(x,y)= h_0{\theta(x^2-d^2)},
\end{equation}
with two walls of height 1\,nm located at $x=\pm d=\pm1.5$\,nm. In
Fig.~\ref{figwellsup} we show the optimum configuration of arm-chair
graphene with supported boundary condition on top of the trench
defined by~Eq.~(\ref{Eq.trench}).

\begin{figure}
\begin{center}
\includegraphics[width=0.9\linewidth]{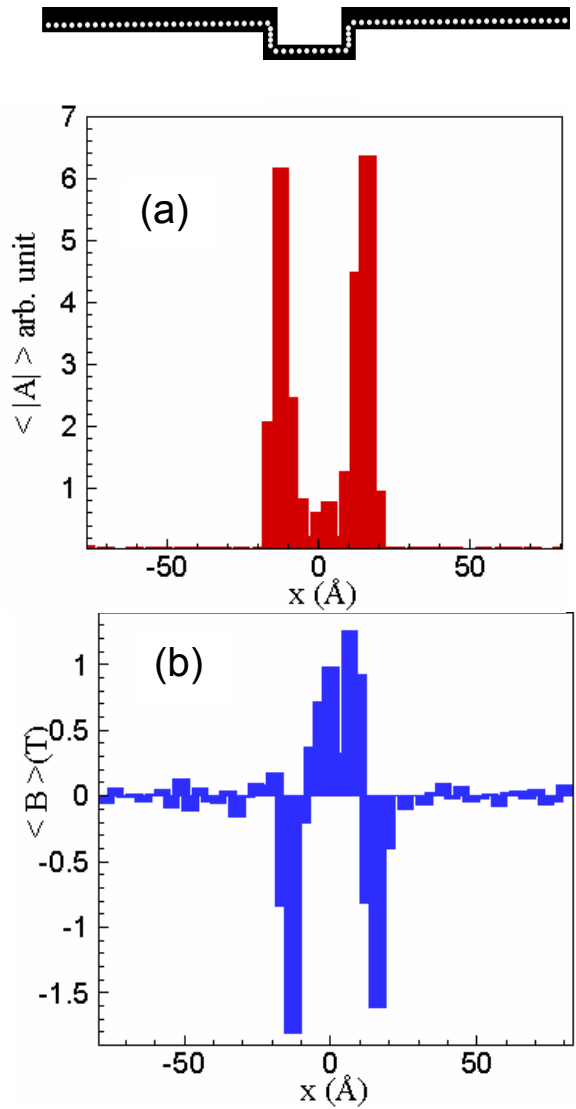}
\caption{(Color online)  The averaged  gauge field (a)
 and the induced pseudo-magnetic field (b) averaged over the $y$-direction for
graphene on top of  a well as shown in Fig.~\ref{figwellsup} which
has been supported from the longitudinal ends while it can freely
move along the $z$-direction. \label{ABwell} }
\end{center}
\end{figure}

%
%\begin{figure}
%\begin{center}
%\includegraphics[width=1\linewidth]{bwellx.eps}
%\caption{(Color online) The induced pseudo-magnetic fields for an
%arm-chair graphene shown in Fig.~\ref{figwellsup} which has been
%supported from the longitudinal ends while it can vibrate in the
%$z$-direction. \label{figwellB} }
%\end{center}
%\end{figure}

The absolute value of the induced gauge field  as obtained from
Eq.~(\ref{gauge}) is averaged over the $y$-direction and is shown in
Fig. \ref{ABwell}(a). All atoms at both sides are stretched toward
the well region which results in a considerable gauge field around
$x\approx \pm d$. Fig.~\ref{ABwell}(b) shows the
$y$-averaged pseudo-magnetic field. %over the $y$-direction, $\langle
%B \rangle$ which is less than 1\,Tesla. However, Fig.~\ref{figwellB}
%shows that the distribution of the induced pseudo-magnetic field,
%$B$ (calculating from Eq.~(\ref{figwellB})), results in a maximum
%$B$ of about 100\,Tesla.
Notice that there is a non-zero $\langle |\textbf{A}| \rangle$ and
$\langle B \rangle$ within the trench which is a consequence of the
bent (non-flat) graphene in the middle region. Notice that the
pseudo-magnetic fields are smaller than those obtained for a step
profile. Indeed supporting GE longitudinally from two ends prevents
  GE to move into the well and consequently there will be  less variations in the
heights. The magnetic filed profile consists of a positive B-barrier
inside the trench and two negative barriers located at the steps.
The total average magnetic filed is also zero in this case.

\subsection{ Barriers}

A barrier in the middle of the substrate is reverse situation of the
previous case. An elongated barrier in the $y$-direction is
parameterized as
\begin{equation}\label{Eq.barrier}
h_S(x,y)=h_0{\theta(x^2-d^2)},
\end{equation}
with two walls at $x=\pm d=\pm$1.5\,nm of height of 1\,nm.

\begin{figure}
\begin{center}
\includegraphics[width=0.98\linewidth]{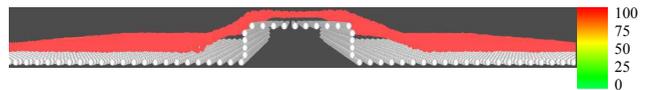}
\caption{(Color online)  The optimum configuration of arm-chair
graphene over an elongated cubic barrier with $|x|<1.5$\,nm where
the zig-zag edges were supported in the $x-y$ plane. The colors
indicate the size of strain. \label{figbarsup} }
\end{center}
\end{figure}

\begin{figure}
\begin{center}
\includegraphics[width=0.9\linewidth]{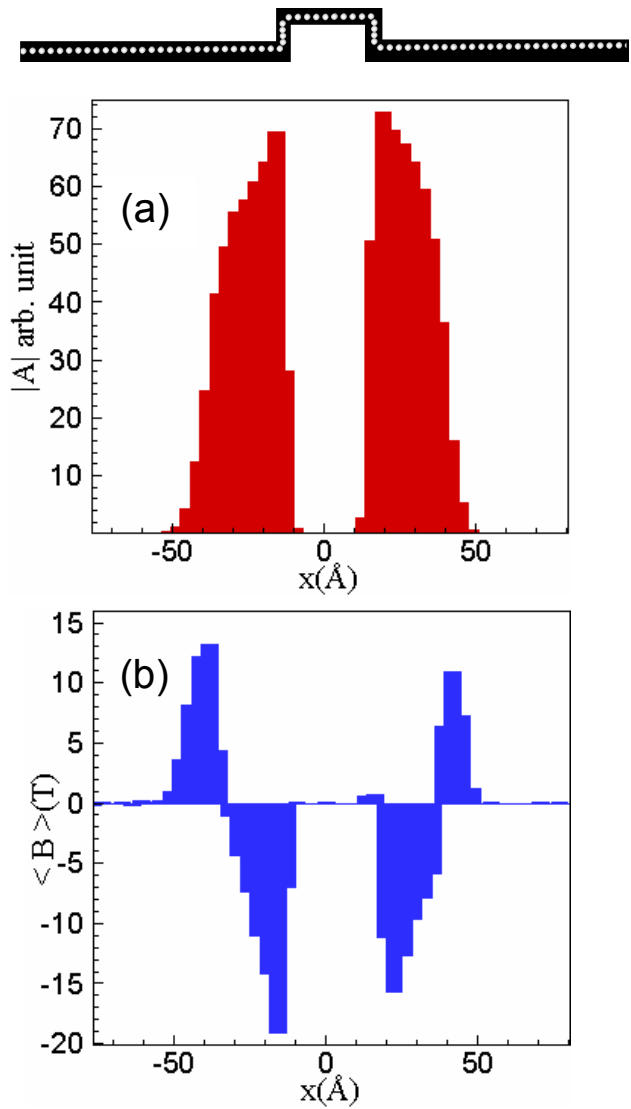}
\caption{(Color online)  The averaged  (a) gauge field in
 and (c) the induced pseudo-magnetic field averaged over the $y$-direction for a
graphene on top of a barrier as shown in Fig.~\ref{figbarsup} which
has been supported from the longitudinal ends while it can freely
move along the $z$-direction.  \label{ABbarier} }
\end{center}
\end{figure}

%\begin{figure}
%\begin{center}
%\includegraphics[width=1\linewidth]{bbarrier.eps}
%\caption{(Color online) The induced pseudo-magnetic fields for
%arm-chair graphene shown in Fig.~\ref{figbarsup} which has been
%supported from the longitudinal ends while it can vibrate along the
%$z$-direction.  \label{figbarB} }
%\end{center}
%\end{figure}

Figure~\ref{figbarsup} shows the optimum configuration of arm-chair
GE in the case of  supported boundary condition over the barrier.

The induced gauge field  as obtained from Eq.~(\ref{gauge}) was
averaged over the $y$-direction and is shown in Fig.
\ref{ABbarier}(a). All atoms at both sides are stretched towards the
barrier region which causes considerable gauge fields around
$x\approx \pm d$. Fig.~\ref{ABbarier}(b) shows the averaged
pseudo-magnetic field over the $y$-direction which is
less than 20\,Tesla. %However as seen from Fig.~\ref{figbarB}, which
%shows the distribution of the induced pseudo-magnetic field, $B$
%(calculating from Eq.~(\ref{Magnetic})), the maximum absolute value
%of $B$ is about 350\,Tesla.
Both gauge and pseudo-magnetic fields are comparable with those
found for the substrate with a single step placed in the middle of
the GE sheet. The main difference is the formation of a
zero-magnetic field channel in the region $|x|<d$. The electrons
will be trapped in this rectangular channel which can be realized in
experiments. On both sides of this magnetic channel there are two
double magnetic barriers of similar shape. Because of thermal
fluctuations the barriers are not identical.

\subsection{ Spherical bubble}
\begin{figure}
\begin{center}
\includegraphics[width=0.98\linewidth]{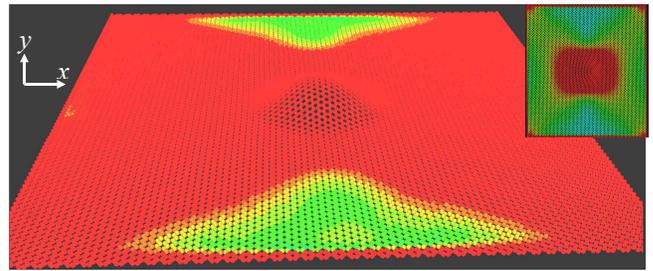}
\caption{(Color online)  The optimum configuration of arm-chair
graphene, with two longitudinal ends  supported in the $x-y$ plane ,
on top of a bubble. The inset shows a different view indicating the
elongation of the deformation of graphene along the $x$-direction,
i.e. arm-chair direction. The colors indicate the size of strain.
\label{figbublsup} }
\end{center}
\end{figure}

The next important deformation of the substrate that has been
realized experimentally~\cite{Bubl1,Bubl2} is a bubble
\begin{equation}
 h_S(x_i,y_i)=\sqrt{R^2-\rho_i^2}+h_1,\label{hbump}
 \end{equation}
where $R$ is the radius of the bubble and $\rho_i^2=x_i^2+y_i^2$ is
the radial distance of the $i^{th}$ atom from the center. In order
to create an uniform discrete atomistic structure for the bubble, we
set $h_1=-R/2$ where $R=$2\,nm. The optimum configuration for the
longitudinally supported graphene over the bubble substrate is shown
in Fig.~\ref{figbublsup}. Due to the supported end GE is elongated
longitudinally along the supported direction, see inset in
Fig.~\ref{figbublsup}.

\begin{figure}
\begin{center}
\includegraphics[width=0.46\linewidth]{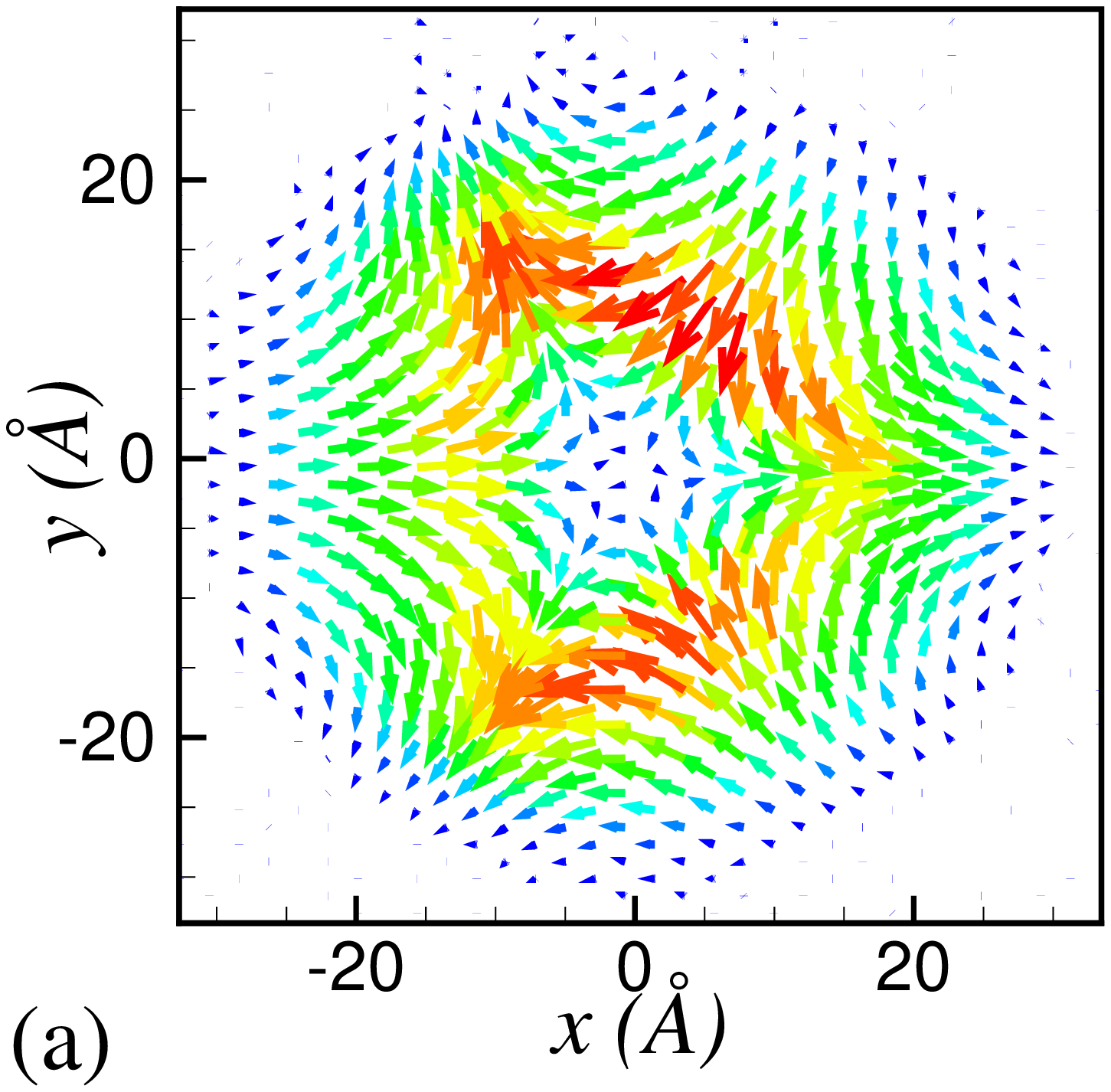}
\includegraphics[width=0.52\linewidth]{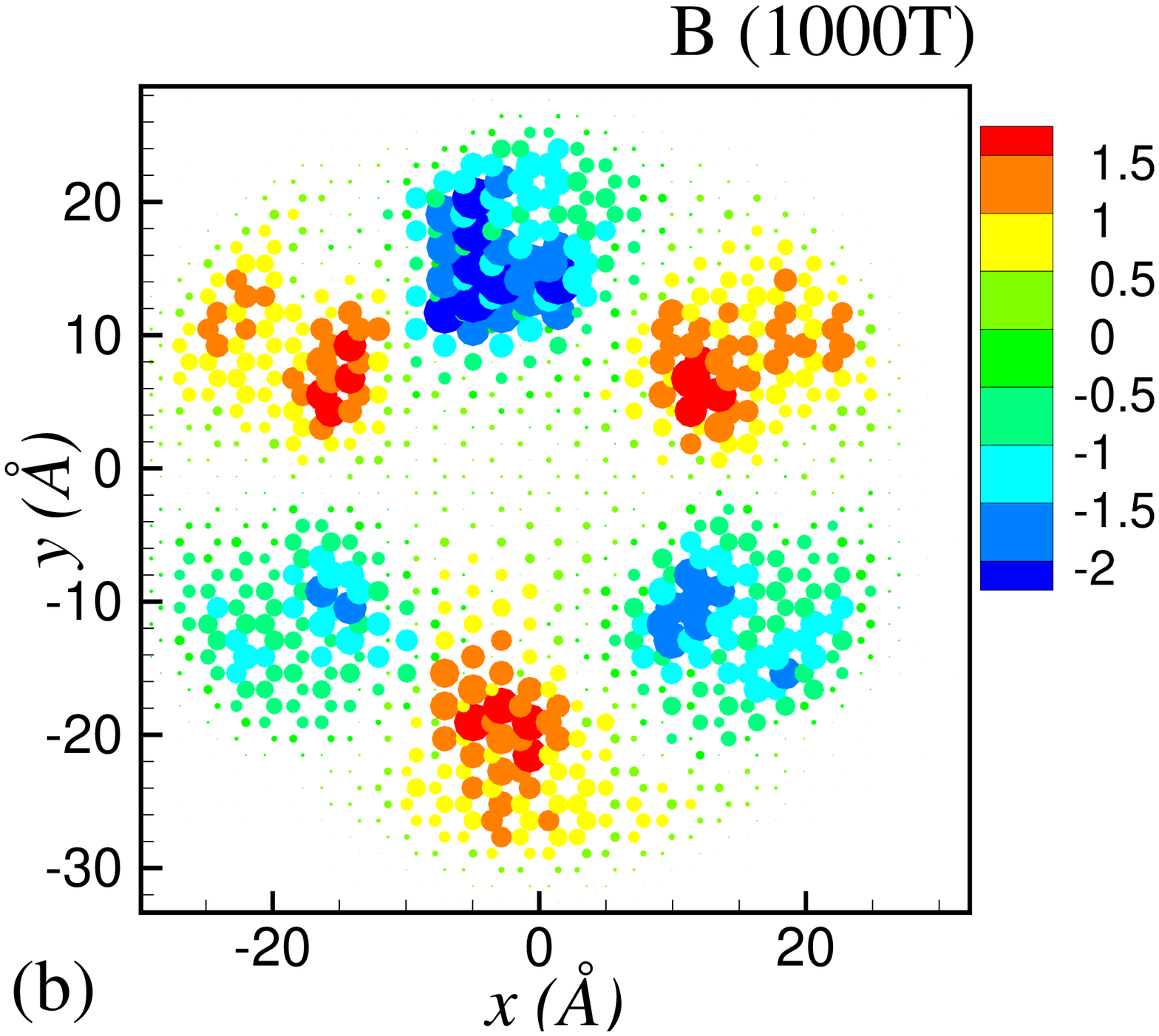}
\caption{(Color online)  (a) Vector plot of the gauge field and (b)
the induced pseudo-magnetic field for arm-chair graphene placed over
a spherical bubble. The obtained deformation is shown in
Fig.~\ref{figbublsup}. Graphene  was supported from the longitudinal
ends while it can vibrate along the $z$-direction. \label{figbubleB}
}
\end{center}
\end{figure}

In Fig. \ref{figbubleB}(a) the  induced gauge field (corresponding
to the induced strain and around the central part) is illustrated by
using Eq.~(\ref{gauge}). Figure~\ref{figbubleB}(a) shows a vector
plot of the induced gauge fields where the length of the vectors and
the colors denote the absolute value of  $\textbf{A}$. The
corresponding magnetic field is depicted in Fig.~\ref{figbubleB}(b).
Notice that both the gauge field and the pseudo magnetic field
exhibits  an approximate six fold
symmetry~\cite{PRBrapid,naturenano}. Becasue of thermal fluctuations
the symmetry is not exact. Notice that there is a little elongation
along the supported direction. We will discuss this symmetry in the
next section. Notice that the induced magnetic fields are larger
than those found for the step, trench and barrier. %Note that the
%fields far from the central part, Gaussian compound, are negligible,
%i.e. $|\textbf{A}|,B\approx0$.

\subsection{Gaussian bump/depression}

The  Gaussian~bump (protrusion)/depression~\cite{neekJPCM,Piniing}
is parameterized as
\begin{equation}
h_S(x_i,y_i)=\pm h_0 \exp(-\rho_i^2/2\gamma^2),\label{hbump}
 \end{equation}
 where $+h_0(-h_0)$(=1\,nm) is the height (depth) of the Gaussian bump (depression)
and ${\rho_i}^2=x_i^2+y_i^2$ is the radial distance of $i^{th}$ atom
and $\gamma=1$\,nm is the variance of the Gaussian.

Since the optimum configuration of  supported graphene over the
Gaussian bump is similar to the one for a spherical bubble, we will
not report them here. For  supported graphene over a Gaussian
depression the optimum configuration is not Gaussian (as was shown
in Ref.~\cite{neekpeet}).

The results of the gauge fields and pseudo magnetic fields
 for a graphene membrane with Gaussian deformation show a clear six fold
symmetry~\cite{PRBrapid,naturenano}. For a Gaussian deformation of
the graphene membrane, Eq.~(\ref{hbump}), using Eq.~(\ref{gauge})
and Eq.~(\ref{Magnetic}) we found
\begin{equation}
\textbf{A}=-\frac{\rho^2
h^2(\rho,\theta)}{2\gamma^4}(cos(2\theta),\sin(2\theta)/2),
\end{equation}
and for the corresponding pseudo-magnetic field
\begin{equation}\label{Bcont}
 \textbf{B}=\nabla \times \textbf{A}=\frac{h^2(\rho,\theta) \rho^2}{\gamma^6}\sin(3\theta)
\end{equation}
where $x=\rho \cos(\theta)$ and  $y=\rho \sin(\theta)$. The well
known six fold symmetry is due to the dependence of $B$ on
$sin(3\theta)$. Our atomistic results for $B$ (see
Fig.~\ref{figbubleB}(b)) are in good agreement with
Eq.~(\ref{Bcont}).

\begin{figure}
\begin{center}
\includegraphics[width=0.45\linewidth]{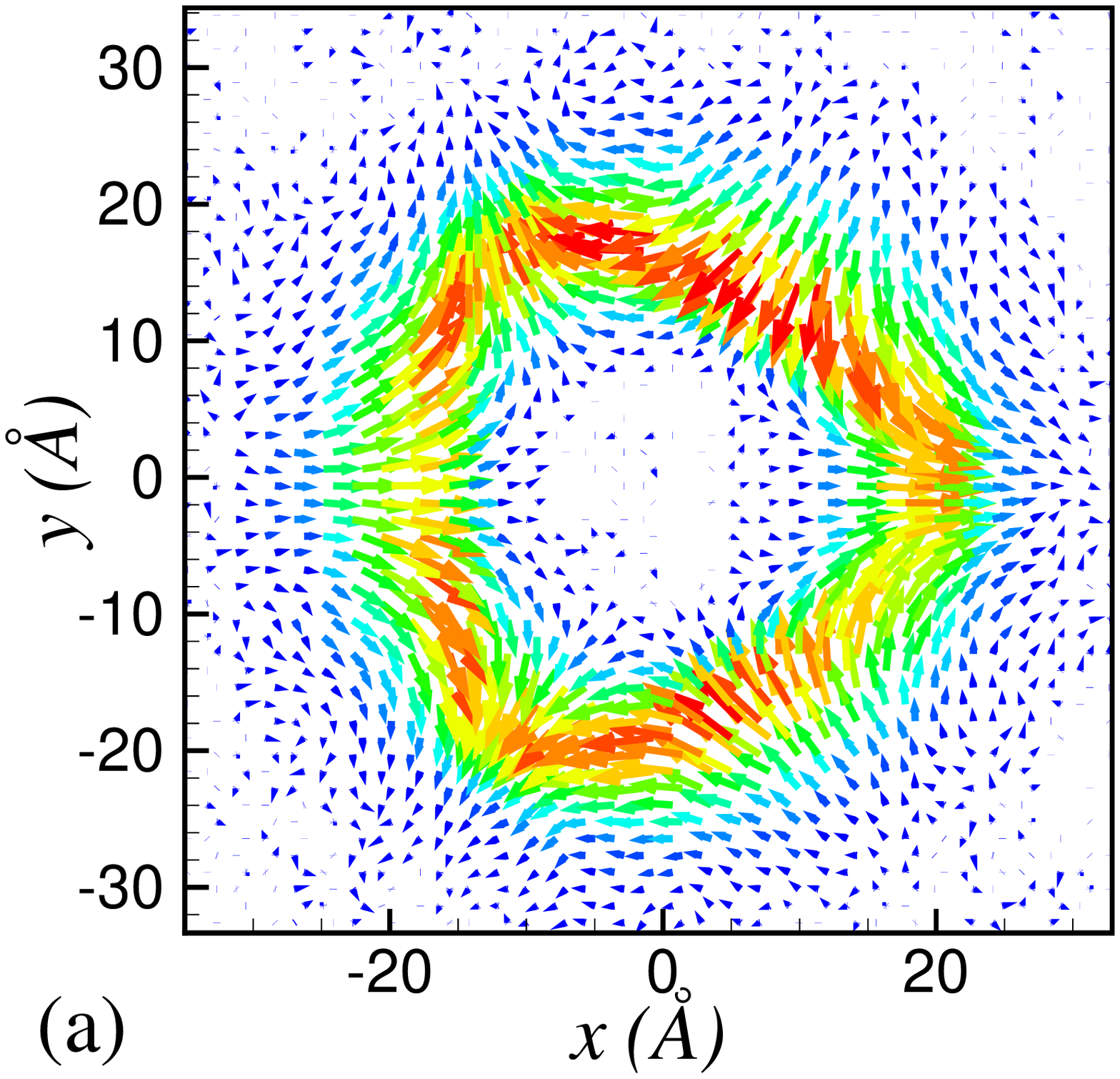}
\includegraphics[width=0.52\linewidth]{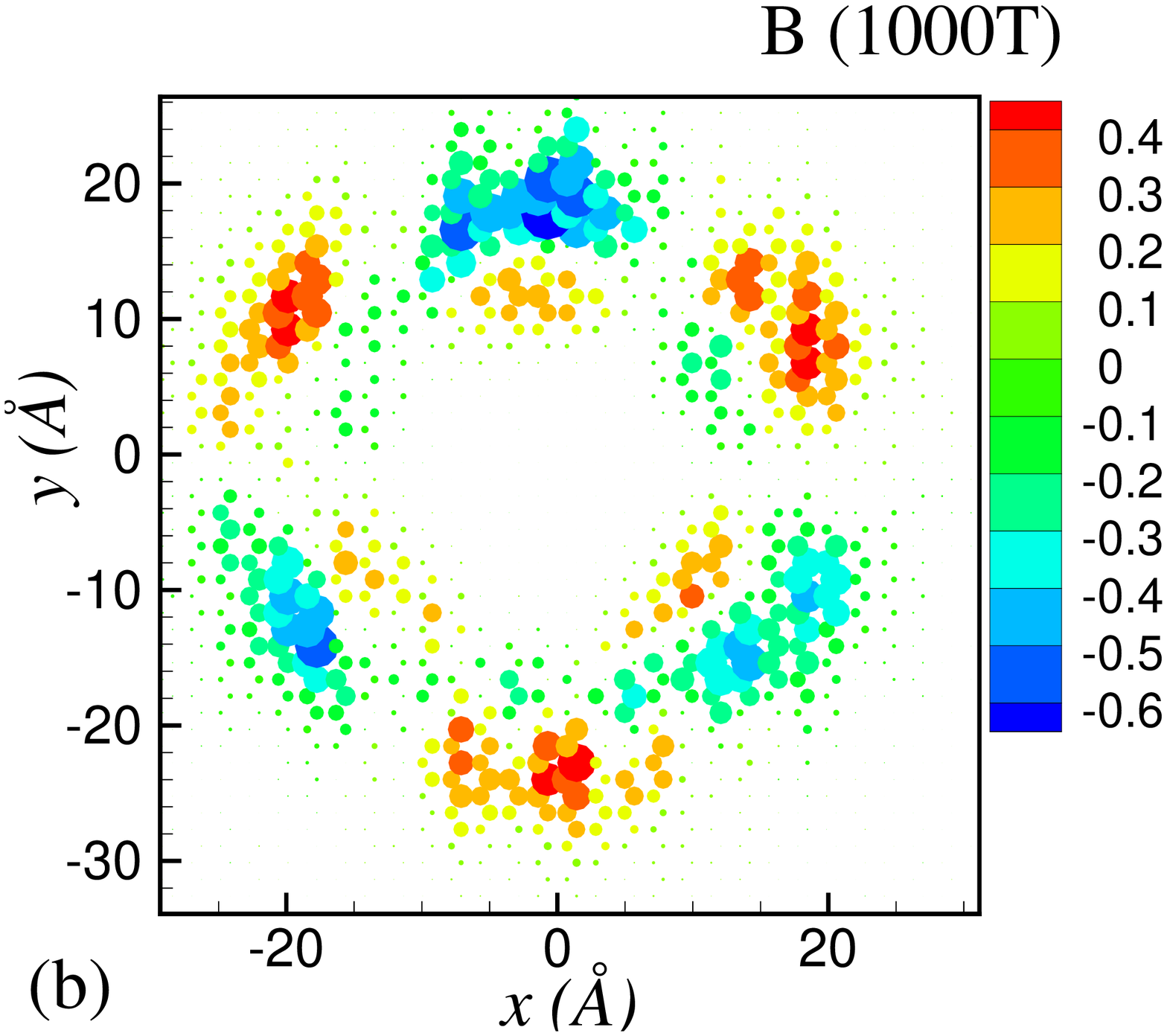}
\caption{(Color online)  (a) Vector plot of the gauge field and (b)
the induced pseudo-magnetic field for arm-chair graphene deposited
over a Gaussian depression. Graphene has been supported from the
longitudinal ends while it can vibrate along $z$-direction.
\label{figdepB} }
\end{center}
\end{figure}

It is surprising that we found a six fold symmetry for deformed
graphene over a Gaussian bump but not for the Gaussian depression.
This is due to the non-Gaussian profile of GE on top of a Gaussian
depression~\cite{neekpeet}. Breaking the radial symmetry of graphene
deformation reduces the symmetry in $\textbf{B}$ (see
Fig.~\ref{figdepB} (b)). This particular symmetry affect also the
energy eigne values and corresponding wave
functions~\cite{PRBrapid}.

In Fig. \ref{figdepB}(a) the induced gauge fields for GE on top of a
Gaussian depression  is shown where the length of the vectors and
the colors denote the absolute value of $\textbf{A}$. Both gauge
fields and pseudo-magnetic fields are smaller than those found for
the supported graphene over the bubble and the Gaussian bump. This
is a consequence of the non-Gaussian profile for the optimum
configurations which yields alteration in six fold symmetry in the
induced pseudo-magnetic field.

\section{Summary}

We investigated systematically the induced pseudo-magnetic field
properties for graphene deposited  on top of different
nanostructured  substrates by using molecular dynamics simulations
at $T$=300\,K. The van der Waals interaction between the substrate
and graphene was modeled by a Lennard-Jones potential. We found that
the induced magnetic filed for graphene on top of a step consists of
two magnetic barriers with different sign, while for a  trench it
forms two narrow magnetic barriers around the trench boundaries and
one with opposite sign within the trench. The one directional
substrate barrier, forms two sets of magnetic-barriers around the
barrier wall. The magnetic filed for the Gaussian depression looses
its six-fold symmetry (due to the non-Gaussian deformation of
graphene) as compared to GE on top of the Gaussian bump/buble. The
strain induced strong pseudo-magnetic fields. Controlling the
pseudo-magnetic field
is possible by controlling  the substrate pattern and the size of the deformation.\\

\pagebreak
 \emph{{\textbf{Acknowledgment}}}. This
work was supported by the Flemish Science Foundation (FWO-Vl) and
ESF EUROCORE program EuroGRAPHENE: CONGRAN.

\end{document}